**Airway cross-sectional area measurement using quantitative endoscopy with AI-enabled depth estimation: A proof-of-concept study**

Amirsajjad Taleban,[1] Jake Luo,[1] Christopher E. Larkee,[2]

B. Tucker Woodson,[3] Ryan Puccia,[3] John S. Rhee,[3] Guilherme J.M. Garcia[2,3] *

1 – University of Wisconsin – Milwaukee, Department of Health Informatics and Administration, Milwaukee, Wisconsin, USA

2 – Marquette University & The Medical College of Wisconsin, Joint Department of Biomedical Engineering, Milwaukee, Wisconsin, USA

3 – Medical College of Wisconsin, Department of Otolaryngology and Communication Sciences, Milwaukee, Wisconsin, USA

Corresponding author:

* Guilherme Garcia, PhD

8701 Watertown Plank Road, Milwaukee, WI, 53226, United States

Phone: (414) 955-4466 / Fax: (414) 955-6568

Email: ggarcia@mcw.edu

ORCID: https://orcid.org/0000-0003-2995-9226

**Word Count (title page to figure legends):** 4,923 words

**Word Count in Abstract:** 190 words

**Number of figures:** 12

**Number of tables:** 1

## ABSTRACT

Endoscopy is widely used in medicine to visualize anatomical structures for diagnosis and treatment evaluation. However, it does not currently allow rapid, quantitative measurement of airway cross-sectional area (CSA), a key metric for diagnosing and managing respiratory pathologies. There is growing interest in AI-enabled quantitative endoscopy using depth estimation, but a fundamental challenge is the need for ground-truth depth data to train algorithms capable of quantifying depth in physical units (e.g., meters). In this study, 60,000 paired virtual endoscopy images and depth maps were created in Blender from 3D airway models built from computed tomography. A U-Net convolutional neural network was trained to predict the depth maps from the endoscopy images. Point clouds were reconstructed from the AI-generated depth maps using the Equisolid fisheye lens model, and airway perimeters were identified by intersecting a plane with the point cloud to estimate CSA. Nasopharyngeal CSA, a key metric in obstructive sleep apnea, was quantified using this AI-enabled approach with a median error of 5.3% relative to the 3D model ground-truth. This pilot study using synthetic data provides strong evidence for the feasibility of developing AI-enabled quantitative endoscopy for airway CSA measurement.

## INTRODUCTION

Endoscopy is widely employed in the evaluation of upper airway (UA) disorders, providing essential visualization for diagnosis and treatment outcome assessment. However, it lacks the capability for rapid quantification of UA diameter or cross-sectional area (CSA). Previous research studies developed methods to estimate the airway CSA from endoscopy using a reference object of known size [3, 9, 12, 14, 25], pattern projection [10, 21, 22, 27], or stereovision endoscopes [11]. However, no quantitative endoscopy technique has been adopted in clinical practice because these methods require specialized endoscopes, or manual tracing of the airway perimeter, which is labor-intensive and error prone. A new paradigm for quantitative endoscopy is the use of artificial intelligence (AI) to estimate depth maps [8], a technology that is currently used in self-driving vehicles. Our team previously demonstrated that CSAs of 3D nasal cavity models built from computed tomography (CT) scans can be accurately estimated from depth maps [8]. In our previous study, the depth maps could be generated because the 3D geometry was known from the CT scans. The goal of this study is to demonstrate the feasibility of training an AI algorithm to predict depth maps and to evaluate the accuracy of airspace CSAs estimated from AI-generated depth maps.

Artificial intelligence has revolutionized the field of computer vision by enabling accurate three-dimensional scene reconstruction from monocular images, a capability that has found widespread applications in autonomous vehicles, robotics, and medical imaging [4, 15, 16, 34]. Neural networks, particularly convolutional architectures, can learn to predict pixel-wise depth information by capturing spatial hierarchies and geometric relationships within two-dimensional images [6]. In the medical domain, encoder-decoder architectures such as U-Net and ResNet-based networks have demonstrated remarkable capability in estimating depth maps from endoscopic images through hierarchical feature extraction and spatial reconstruction [19, 32, 36]. Early approaches utilized fully convolutional networks combined with conditional random fields to handle the unique challenges of endoscopic scenes, such as specular reflections and tissue deformations [19]. More recent supervised learning methods have employed ResNet-based architectures with skip connections and residual learning to preserve fine-grained spatial details essential for accurate depth prediction [31, 33, 35]. These supervised approaches typically require ground truth depth maps for training, which can be obtained through structured light scanning, stereo vision systems, or synthetic data generation using 3D modeling software [23, 24, 39]. Alternatively, self-supervised learning methods have emerged as promising approaches that learn depth estimation without explicit ground truth labels by exploiting geometric constraints such as photometric consistency between stereo pairs [38] or temporal coherence in video sequences [40]. While these self-supervised techniques offer advantages in terms of reduced annotation requirements, semi-supervised methods trained on high-quality synthetic data have demonstrated superior accuracy and robustness, particularly for clinical applications where precision is paramount [28]. Building on this foundation, our study employs a U-Net architecture trained on synthetic endoscopic images and depth maps to develop a clinically feasible depth estimation system. Recent nasal endoscopy studies show feasible monocular 3D reconstruction with NeRF and related pipelines [13, 26]. Yet depth estimation for the broader upper airway remains underdeveloped, with scarce datasets and limited clinical validation [13, 26].

This proof-of-concept study employs synthetic data—specifically, virtual endoscopy images and depth maps—to demonstrate the feasibility of training an AI algorithm for depth estimation from endoscopic images. Virtual endoscopy images of the nasal cavity and nasopharynx were generated in Blender, an open-source 3D modeling and animation software, using 3D anatomical models reconstructed from CT scans. Each virtual endoscopy image was paired with a corresponding depth map—a grayscale image in which pixel intensity represents the distance from the endoscope camera to the nasal wall. A dataset comprising 60,000 image pairs served as ground truth for training the AI algorithm. The trained AI model was integrated into an in-house software application designed to predict depth maps from endoscopic images, identify the airway perimeter, and compute the cross-sectional area at specified distances from the camera. Our results strongly support the feasibility of developing a quantitative endoscopy technology based on AI-driven depth estimation. By enabling rapid, point-of-care quantification of airway dimensions, this technology has potential applications in the surgical planning of various obstructive upper airway disorders.

## METHODS

*Creation of Virtual Endoscopy Images and Depth Maps*

This study was performed according to the principles of the Declaration of Helsinki. Approval was granted by the Institutional Review Board (IRB) of the Medical College of Wisconsin. Informed consent was obtained from all subjects. A set of 30 patient-specific models of the nasal cavity was selected from a database of 3-dimensional models built from CT scans of healthy subjects [2] and patients with nasal airway obstruction [7, 29]. This set of 30 models was previously selected to represent a wide range of nasal valve geometries spanning from narrow to wide [8]. The virtual endoscopy images and depth maps were created in Blender software using a camera path that started at the nostril and ended at the nasopharynx (**Figure 1A**). The camera path was an airflow streamline originating at the nostril centroid and computed by running a computational fluid dynamics simulation of gentle inhalation [2]. A total of 1,000 pairs of images (virtual endoscopy + depth map) were created for each of the 60 unilateral nasal cavities (left and right sides of 30 subjects), thus providing a dataset of 60,000 pairs of images. All images had 720 x 720 pixels. The virtual endoscopy images were created using an Equisolid fisheye lens model with a 2.5-mm focal length, 125$^{o}$ field of view, and 3.6-mm square sensor. The depth map is 2 x 2 matrix that can be visualized as a grayscale image, where the pixel value corresponds to the distance $d$ from the endoscope camera to the nasal wall (**Figure 1B**). The virtual endoscopy images were exported from Blender as .PNG files, while the depth maps were exported as .EXR files (**Figures 1C,1D**).

*AI Methods*

The color virtual endoscopy images were converted into single-channel grayscale .PNG files for the AI training. Preprocessing involved clamping outlier depth values to suppress light reflection artifacts, followed by normalization to (-0.5, 0.5) to center the depth values around zero, which facilitates gradient flow and accelerates convergence during training. We split the data on a patient-wise basis to prevent any leakage between training and testing. The training set comprised of 48 nasal cavities with 48,000 paired images, and the validation set comprised of 12 nasal cavities with 12,000 paired images, totaling 60,000 paired images. For depth estimation, we employed a U-Net (**Figure 2**) with encoder–decoder symmetry and skip connections. The encoder comprised convolutional blocks followed by max pooling, doubling feature channels at each stage, and terminating in a bottleneck. The decoder mirrored this structure with transpose convolutions for up sampling, skip connections to the encoder outputs, and convolutional refinement, halving feature channels at each stage. The output layer generated a 720 x 720 single-channel depth map. We trained the models for 80 epochs, using mean squared error to exploit the pixel alignment for precise regression, and optimized them with Adam (learning rate = $1\times10^{-4}$). The mean squared error (MSE) was defined as

$$MSE = \frac{1}{N \times 720^2} \sum_{k=1}^{N} \sum_{i=1}^{720} \sum_{j=1}^{720} \left( {d_{ij}}^{(k)} - {\hat{d}_{ij}}^{(k)} \right)^2$$

where $N$ is the number of samples ($N =$ 48,000 for training and $N =$ 12,000 for testing), $d_{ij}$ is the ground-truth depth at pixel position $(i, j)$, and $\hat{d}_{ij}$ is the depth predicted by the AI model. Because the input and output were strictly pixel-wise aligned, with each pixel in the image corresponding directly to its depth value, this alignment simplified the model's learning process and made training easier.

*Estimation of Airway CSA From Depth Maps*

The airway perimeter at a distance $d^*$ from the endoscope camera can be readily obtained from the depth map by selecting pixels with $d > d^*$. In our previous study [8], the airway cross-sectional area ($A$) in units of cm$^2$ was computed via the equation $A = kN_{pixels}$, where $N_{pixels}$ is the number of pixels inside the airway perimeter. The conversion factor $k$ (units of cm$^2$/pixel) is a function of the distance $d$; it was calibrated based on virtual endoscopy images and depth maps of 10 cylindrical tubes of known diameters [8]. This approach had the inconvenience that the CSA was estimated perpendicular to the endoscope camera, thus the camera had to be aligned with the airway axis to measure the airway CSA. In this study, we adopt an alternative approach, namely the depth map is used to compute a point cloud corresponding to the 3D geometry of the nasal cavity, which allows computing the airway CSA perpendicular to the airway axis independent of the camera orientation.

The following steps are performed to compute the point $P = (X, Y, Z)$ in 3D space corresponding to a pixel $p = (x, y)$ in the depth map. First, we compute the polar coordinates of pixel $p = (r, \phi)$ in the 2D image, namely $r = \sqrt{x^2 + y^2}$ and $\phi = \arctan(y, x)$. Here $x$, $y$, and $r$ are measured in numbers of pixels from the image center. Next, we apply the Equisolid fisheye lens model to compute the zenith angle $\theta$ between the camera axis and a vector from the camera to point $P$, namely $\theta = 2\arcsin\left(\frac{c\,r}{2f}\right)$, where $f$ = 2.5 mm is the focal length of the fisheye lens. The constant $c$ is a unit conversion factor accounting for the fact that the 720-pixel endoscopic image is generated by a 3.6-mm square sensor, so that the pixel size is $c$ = (3.6 mm)/(720 pixels) = 0.005 mm/pixel. The depth $d$, zenith angle $\theta$, and polar angle $\phi$ are the spherical coordinates of point $P = (X, Y, Z) = (d, \theta, \phi)$. The cartesian coordinates are given by

$$X = d \sin\theta \cos\phi$$

$$Y = d \sin\theta \sin\phi$$

$$Z = d \cos\theta$$

A software application was developed in Python to display a virtual endoscopy image, estimate the depth map using the AI algorithm, and reconstruct the point cloud from the depth map (**Figure 3**). The application included horizontal sliders that allowed users to define a plane at a given distance from the camera and rotate the plane as needed until the plane was perpendicular to the point cloud at the target location (**Figure 3; Supplementary Video 1**). The airway perimeter was defined by the intersection of the plane with the point cloud (**Figure 3**). The points along the airway perimeter were organized sequentially (**Figure 4**) and the cross-sectional was computed from the contour integral $A = \oint v\, du = |\sum_i v_i\, \Delta u_i|$, where $(u_i, v_i)$ are the coordinates of the perimeter points and $\Delta u_i = u_{i+1} - u_i$ is the horizontal distance between neighboring points.

*Accuracy of CSAs Estimated From Depth Maps*

A potential clinical application of quantitative endoscopy is measuring pharyngeal CSA in patients with obstructive sleep apnea during drug induced sleep endoscopy [9, 14]. To evaluate the accuracy of the CSA estimation, we measured CSA at the distal end of the nasopharynx, which for most patients corresponded to the minimum nasopharyngeal CSA due to the curvature of the soft palate (**Figure 5**). This anatomical landmark was readily identifiable in the 3D models and therefore provided a reliable ground-truth CSA for comparison with depth-map-derived estimates. Among the 30 patient-specific models, 19 included an extension tube (**Figure 5A**), 4 lacked an extension tube (**Figure 5B**), and 7 had an idealized nasopharynx generated from CT scans that did not capture this region [1] (**Figure 5C**). The latter models were excluded from the CSA accuracy analysis because the nasopharyngeal anatomy was not anatomically accurate. For each remaining model, a new pair of images (virtual endoscopy + depth maps) was created with the endoscope camera positioned at the center of the nasopharynx, yielding 1 pair of images per model. This dataset was entirely separate from the 60,000 image pairs used to train the AI algorithm. Using independent sets of images for training and testing is critical to objectively assess the algorithm's generalizability and performance on unseen data. The AI algorithm performed poorly in the four models without an extension tube, preventing accurate delineation of the airway perimeter at the measurement site (**Figure 6**). In these cases, light reflections on the non-anatomical distal wall (corresponding to the distal boundary of the CT segmentation) compromised the accuracy of the AI-generated depth maps (**Figure 6**), likely because the training data contained few examples in which the distal wall was visible. One additional model with an extension tube was excluded because a prominent soft-palate hump partially obstructed the measurement site, preventing complete delineation of the airway perimeter (**Figure 7**). Consequently, CSA-estimation accuracy was evaluated in the remaining 18 models that included an extension tube (**Figures 5A and 8**).

The absolute error (units of $cm^2$) of the CSA estimation was computed via the equation

$$Absolute\ Error = \left|A_{3D} - A_{Depth\ Map}\right|$$

where $A_{3D}$ is the ground truth CSA in the 3D model and $A_{Depth\ Map}$ is the CSA estimated from the depth map. The percent error (units of %) was computed via the equation

$$Percent\ Error = 100\frac{|A_{3D} - A_{Depth\ Map}|}{A_{3D}}$$

The error was quantified for both the CSAs estimated from the ground truth depth maps and from the CSAs estimated from the AI-generated depth maps. This allowed us to distinguish the impact of errors in the AI-generated depth maps from potential inaccuracies in our methods to estimate CSAs from ground truth depth maps.

*Statistical Analysis*

The Mann-Whitney test was used to test the hypothesis that two measurements were statistically different at the significance level $P < 0.05$. The Pearson correlation coefficient was used to assess the strength of the relationship between the CSAs estimated from depth maps and those obtained from the 3D models. A Bland–Altman analysis was performed to evaluate the agreement between the CSA measurements derived from depth maps and those from the 3D models.

## RESULTS

*AI model training*

The AI model was trained for 80 epochs, with the mean squared error (MSE) in the validation set rapidly converging to approximately 0.02 within the first 15 epochs (**Figure 9**). The optimal model was selected at epoch 24, achieving the lowest MSE of 0.014 in the validation set. The small difference between the MSE in the training and validation sets indicates strong generalization to unseen synthetic scenes and minimal overfitting. **Figure 10** illustrates representative depth map predictions from four anatomical regions. The AI-generated depth maps closely match the ground-truth depth maps across the nasal valve, turbinate region, nasal choana, and nasopharynx. The model accurately captured depth gradients in geometrically complex regions, such as the curved turbinate surfaces and the transitional anatomy of the nasal choana. It also preserved smooth depth values in flatter regions while appropriately reproducing sharp depth discontinuities at tissue boundaries.

*Accuracy of CSAs Estimated From Depth Maps*

The nasopharyngeal airspace CSA of the 3D models was on average 1.36 ± 0.71 $cm^2$ (range: 0.38-2.40 $cm^2$) in the 18 models selected for testing. Excellent agreement was observed for the nasopharyngeal airspace CSAs measured from depth maps relative to the known areas in the 3D models. The CSAs estimated from the ground-truth depth maps had a median absolute error of 0.01 $cm^2$ and a median percent error of 0.6%. The CSAs estimated from the AI-generated depth maps had a median absolute error of 0.06 $cm^2$ and a median percent error of 5.3% (**Table 1**). The percent error was < 2% in 16 out of 18 models when using the ground-truth depth maps (**Figure 11A**). This high level of accuracy validates our method for identifying the airway perimeter from the point cloud generated from the depth maps (**Figures 3 and 4**). Nasopharyngeal airspace CSAs estimated from the AI-generated depth maps had a modestly larger error, but the percent error was ≤ 22% in all cases and < 10% in 13 out of 18 models (**Figure 11A**). CSAs estimated from depth maps were not statistically different from the known values in the 3D models for either the ground-truth depth maps ($p = 0.96$) or the AI-generated depth maps ($p = 0.99$). There was a strong correlation between the CSAs estimated from depth maps and the known values, with a Pearson $r = 0.998$ ($p < 0.0001$) for the ground-truth depth maps and $r = 0.987$ ($p < 0.0001$) for the AI-generated depth maps (**Figure 11B**). The Bland-Altman plot demonstrated negligible bias and no systematic error, with good agreement between CSAs estimated from depth maps and known values across both small and large measurements (**Figure 12**).

## DISCUSSION

Quantitative endoscopy has many potential applications in clinical medicine, including assessing the severity of nasal valve stenosis in patients with nasal airway obstruction, grading tracheal narrowing in subglottic stenosis, and quantifying the size of the pharyngeal airspace in patients with obstructive sleep apnea during drug-induced sleep endoscopy. Prior studies have explored several quantitative endoscopy techniques, such as manual boundary outlining, laser-projected grids for surface triangulation, and stereovision endoscopes equipped with dual cameras. While these approaches have shown promising preliminary results, none has achieved widespread clinical adoption because they are labor-intensive or require equipment not routinely available in standard clinical practice.

There is growing enthusiasm for quantitative endoscopy techniques that extract geometric information from AI-generated depth maps. These methods operate on standard monocular endoscopy, thus they hold strong potential for rapid adoption in clinics and operating rooms, where monocular endoscopes are already ubiquitous. Besides eliminating the need for radiographic studies that expose patients to ionizing radiation, quantitative endoscopy enables repeated image and data capture across a dynamic continuum—for example, during different phases of the breathing cycle (inhalation vs. exhalation), body postures (upright vs. supine), neck positions (flexion vs. extension), and varying degrees of mucosal engorgement (congested vs. decongested). A fundamental challenge, however, is the need for ground-truth depth data to train AI algorithms – that is, the X, Y, and Z coordinates of surface points expressed in physical units (e.g., meters). Such data are difficult to obtain in vivo, as endoscopes lack integrated depth-sensing hardware. Consequently, many prior studies have focused on training AI models to predict scale-ambiguous depth maps, which capture relative surface shape but not absolute dimensions. These approaches are appropriate for applications where metric accuracy is not required, such as detecting unsurveyed mucosal regions in colonoscopy [18].

Despite the growing interest in AI-driven depth estimation for endoscopy, relatively few studies have explored generating depth maps of airway geometries in physical units. Li et al. (2025) developed a method for absolute depth estimation from monocular endoscopy, generating depth maps in Unity using 3D reconstructions created from CT scans of airway phantoms [17]. In the broader field of endoscopic depth estimation, Toussaint et al. (2025) took a different approach, rigorously evaluating foundation models for metric depth estimation through their EndoSynth dataset, pairing synthetic endoscopic images with perfect depth maps derived from computer graphics rendering [30]. Meanwhile, Martyniak et al. (2025) pushed the boundaries of realism by generating photorealistic endoscopic data through surgical simulation combined with diffusion models, yielding training pairs with absolute depth information from physics-based tissue deformation models [20]. These studies consistently reported good agreement between their AI-generated depth maps and ground truth measurements. However, despite their technical sophistication, they all stopped short of the critical clinical application, which is estimating airway cross-sectional area.

Our previous study [8] demonstrated that Blender can be used to generate realistic virtual endoscopy images and corresponding ground-truth depth maps derived from CT-based 3D airway models. Importantly, we showed that these synthetic depth maps enabled accurate estimation of airway CSA [8]. The present study extends this work by showing that synthetic data can be used to train an AI model, and that the resulting AI-generated depth maps can accurately reproduce airway CSAs. Specifically, we found a median percent error of 5.3% (**Table 1**) when comparing measurements derived from the AI-generated depth maps with the 3D models. To our knowledge, this is the first demonstration that AI-generated depth maps can be used to obtain accurate airway CSA measurements.

The field of endoscopic depth estimation has witnessed remarkable methodological diversity, with researchers developing increasingly complex architectures to tackle the unique challenges of endoscopic imaging. Some investigators have embraced complex dual-branch designs that adaptively process local and global features [5], while others have integrated advanced attention mechanisms to handle illumination variations and tissue inhomogeneities [37]. Amidst this rich landscape of methodological innovation, our study deliberately chose a more straightforward U-Net architecture, which proved remarkably effective for upper airway depth estimation. This approach provides computational efficiency, requiring only modest GPU resources on consumer hardware. Our use of synthetic training data eliminates the dependency on large annotated clinical datasets and reduces the need for complex domain adaptation pipelines that typically bridge the gap between pre-trained models and specific medical imaging tasks. While we acknowledge the sophistication and potential advantages of alternative approaches, our method offers practical advantages for clinical translation.

Several limitations of this study should be acknowledged. First, the work is purely computational. Future studies are needed to determine whether airway CSA can be accurately estimated from clinical endoscopy. Second, the sample size used to evaluate the accuracy of CSA estimates from AI-generated depth maps was relatively small (18 geometries), and additional research with larger datasets is required to further validate the method. Third, depth maps in this study were estimated from a single endoscopy image. Although this approach performed well on our dataset of computer-generated endoscopy images, developing a similarly robust method for clinical endoscopy will likely be more challenging. Future studies should explore AI approaches that leverage the temporal component of endoscopy—such as using multiple consecutive frames from a video—to estimate depth. Methods that incorporate multiple frames may yield more accurate depth estimation than those relying on a single image.

## CONCLUSIONS

Endoscopy currently lacks methods for rapid, quantitative assessment of airway cross-sectional area (CSA). In this proof-of-concept study, we used Blender to generate a synthetic dataset of 60,000 paired virtual endoscopy images and depth maps derived from 3D models of the nasal cavity and nasopharynx. A U-Net convolutional neural network trained on this dataset accurately reproduced ground-truth depth maps across diverse anatomical regions. Building on these outputs, we developed a novel technique to estimate airway CSA by applying the Equisolid fisheye lens model to AI-generated depth maps to reconstruct 3D point clouds, enabling perimeter detection and CSA computation at target locations. When applied to 18 nasopharyngeal geometries, this method yielded a median percent error of 5.3% relative to known CSA values. This pilot study, based entirely on synthetic data, demonstrates the feasibility of AI-enabled quantitative endoscopy for airway CSA measurement. Future work should validate this approach using clinical endoscopy data and evaluate its performance across broader patient populations and anatomical variability.

## Declarations

### Funding

This project was funded in part by the Advancing a Healthier Wisconsin Endowment (project number GR-07310), in part by the Research Innovation Fund, a philanthropic donation to the Joint Department of Biomedical Engineering at Marquette University and the Medical College of Wisconsin, and in part by NIH/National Institute of Biomedical Imaging and Bioengineering (R01EB009557). The sponsors had no role in the study design, data collection, analysis, interpretation, manuscript writing, or the decision to submit the manuscript for publication.

### Competing Interests

The authors have no relevant financial or non-financial interests to disclose.

### Ethics approval

This study was performed according to the principles of the Declaration of Helsinki. Approval was granted by the Institutional Review Board (IRB) of the Medical College of Wisconsin.

### Consent to participate

Informed consent was obtained from all subjects.

**Author contributions**

**Amirsajjad Taleban:** Investigation; Methodology; Software; Validation; Writing – original draft; Writing – review and editing; **Jake Luo:** Funding acquisition; Investigation; Methodology; Software; Project administration; Supervision; Validation; Writing – review and editing; **Christopher Larkee:** Funding acquisition; Conceptualization; Investigation; Methodology; Validation; Software; Visualization; Writing – review and editing; **Tucker Woodson:** Funding acquisition; Writing – review and editing; **Ryan Puccia:** Funding acquisition; Writing – review and editing; **John Rhee:** Conceptualization; Funding acquisition; Writing – review and editing; **Guilherme Garcia:** Conceptualization; Formal analysis; Funding acquisition; Investigation; Methodology; Project administration; Supervision; Validation; Writing – original draft; Writing – review and editing.

## TABLE

**Table 1** – Median absolute error and percent error for the nasopharyngeal airspace cross-sectional area estimated from ground-truth and AI-generated depth maps relative to the known values measured in the 3D models.

| | Median Absolute Error | Median Percent Error |
|---|---|---|
| Ground-truth depth map | 0.01 $cm^2$ | 0.6% |
| AI-generated depth map | 0.06 $cm^2$ | 5.3% |

## FIGURES

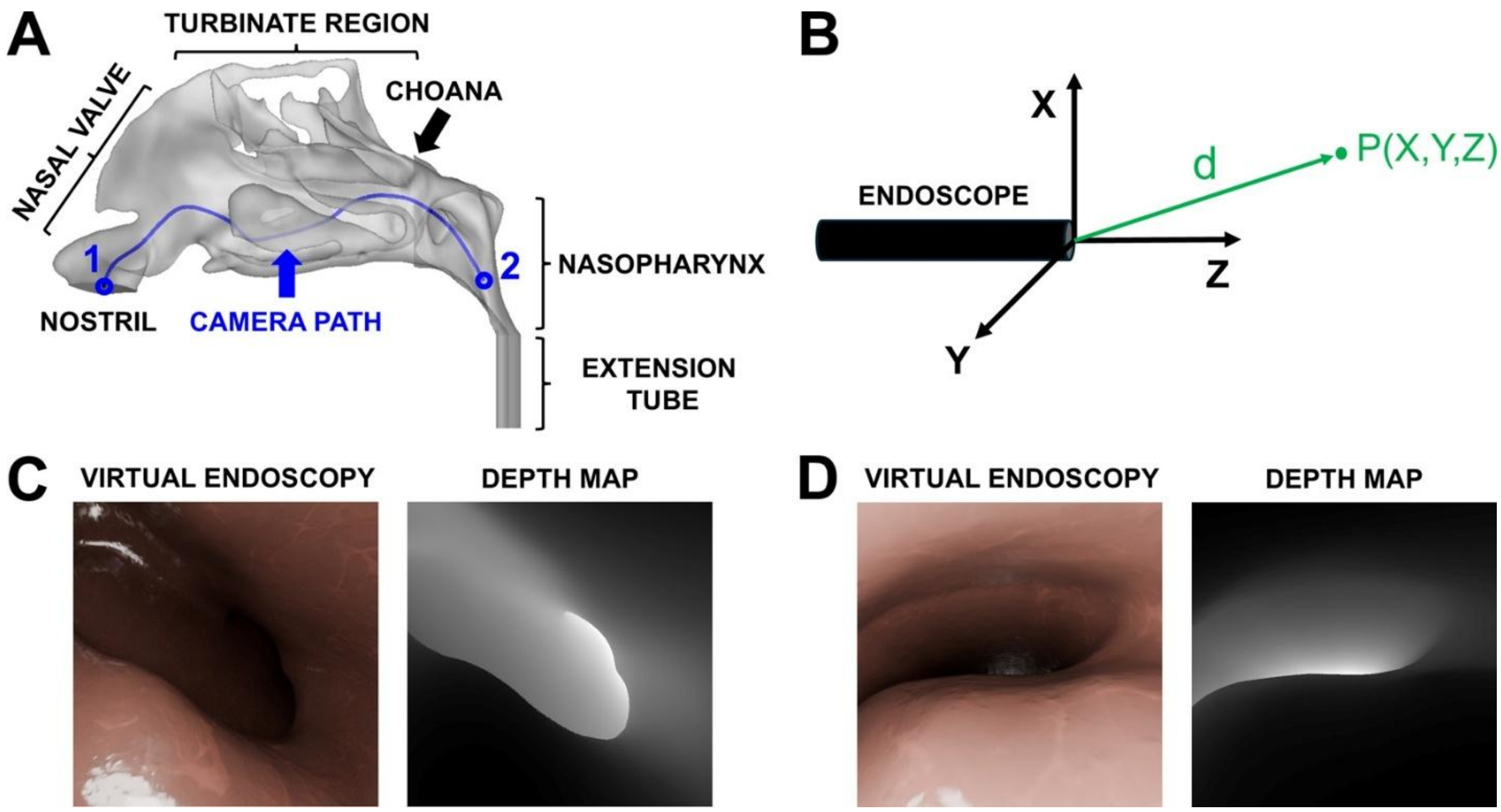

**Fig. 1** Creation of virtual endoscopy and depth map images. (A) Patient-specific model of the nasal cavity showing the camera path. (B) Schematic of the coordinate system positioned on the endoscope camera. Point P(X,Y,Z) on the wall of the nasal cavity is located at a distance $d$ from the endoscope camera. (C) Virtual endoscopy and depth map images with the endoscope camera at the nostril (point 1 in panel A). (D) Virtual endoscopy and depth map images with the endoscope camera at the nasopharynx (point 2 in panel A).

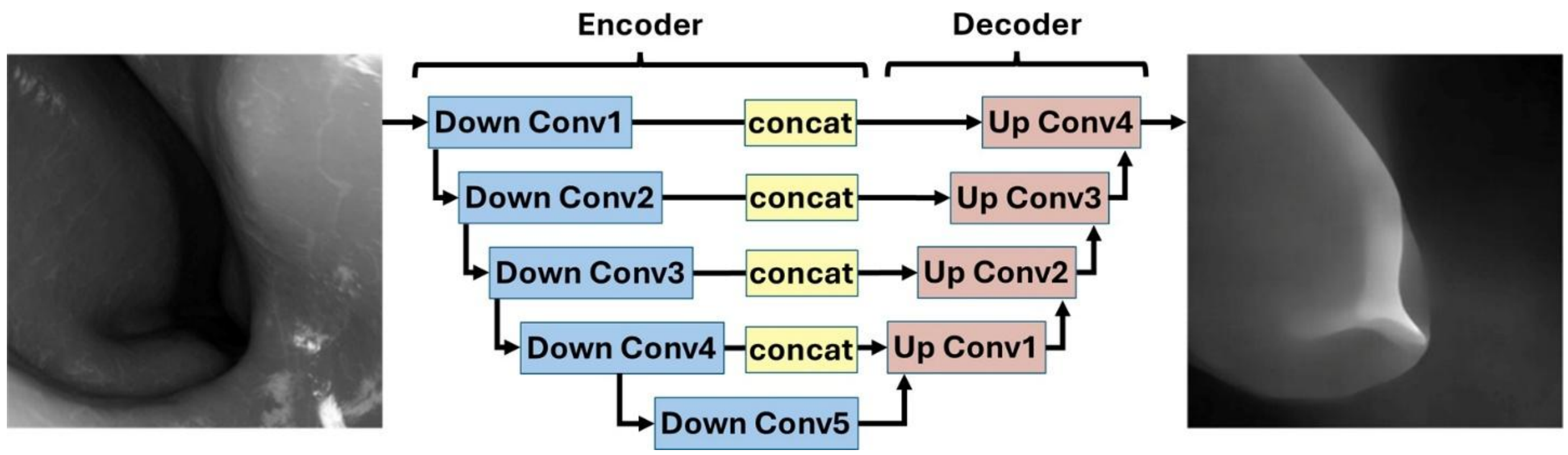

**Fig. 2** U-Net architecture for monocular depth estimation. The encoder (blue boxes) progressively extracts features through five downsampling blocks, followed by four decoder blocks (pink boxes) that reconstruct the depth map using skip connections (yellow concat operations). Input grayscale endoscopic image (left) is transformed to a continuous depth map (right).

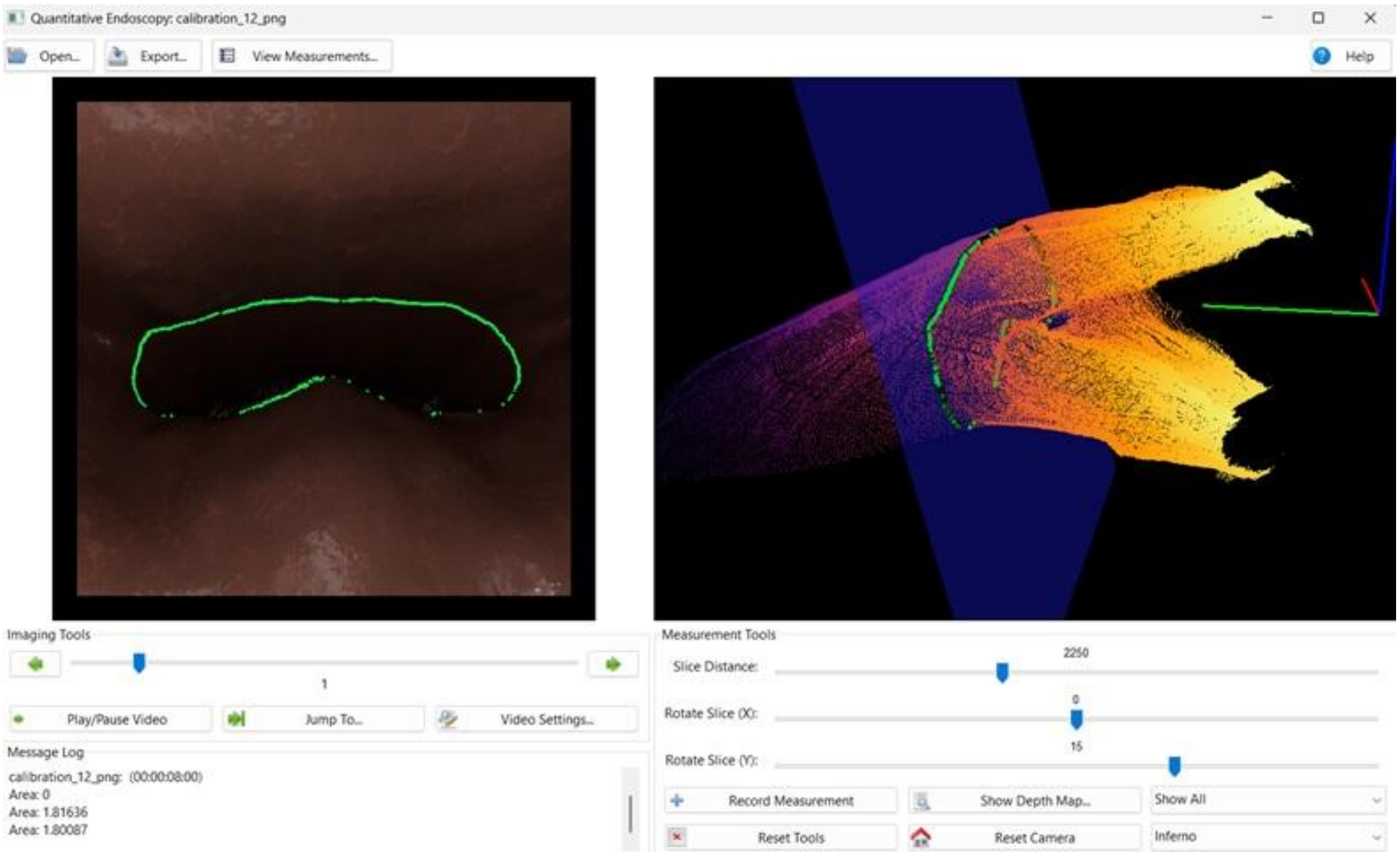

**Fig. 3** Software application used to measure airway cross-sectional areas. The AI algorithm processes the virtual endoscopy image (left), producing a depth map (not shown) that is used to reconstruct the point cloud (right). The first horizontal bar allows users to select a specific distance from the camera, while the second and third bars allow users to rotate the planar intersection and identify the airway perimeter (green points). The airway cross-sectional area is computed from this perimeter and displayed on the text box (bottom left).

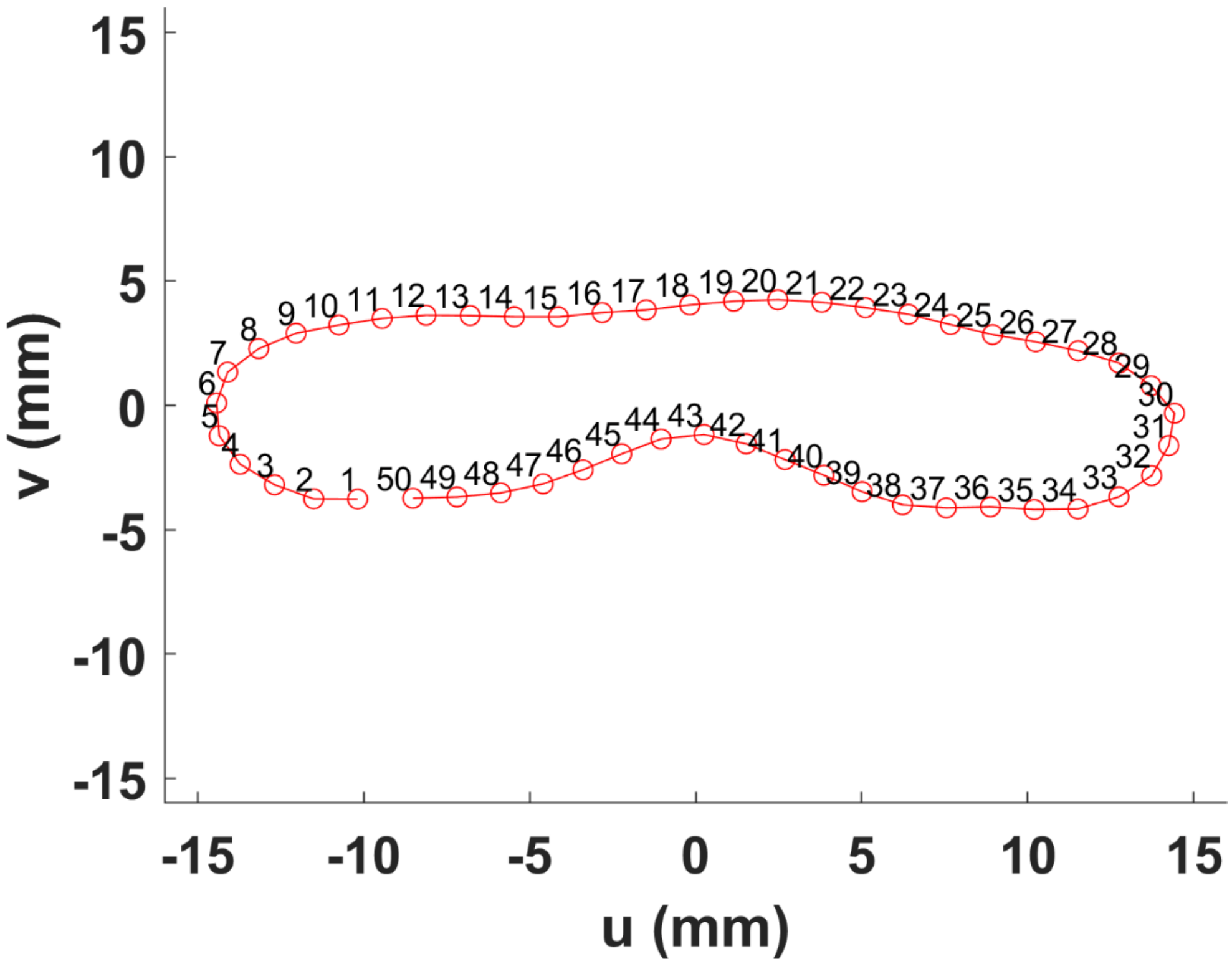

**Fig. 4** Airway cross-sectional is computed from the coordinates $(u, v)$ of points along the airway perimeter on a plane perpendicular to the airway using the integral $A = \oint v \, du$.

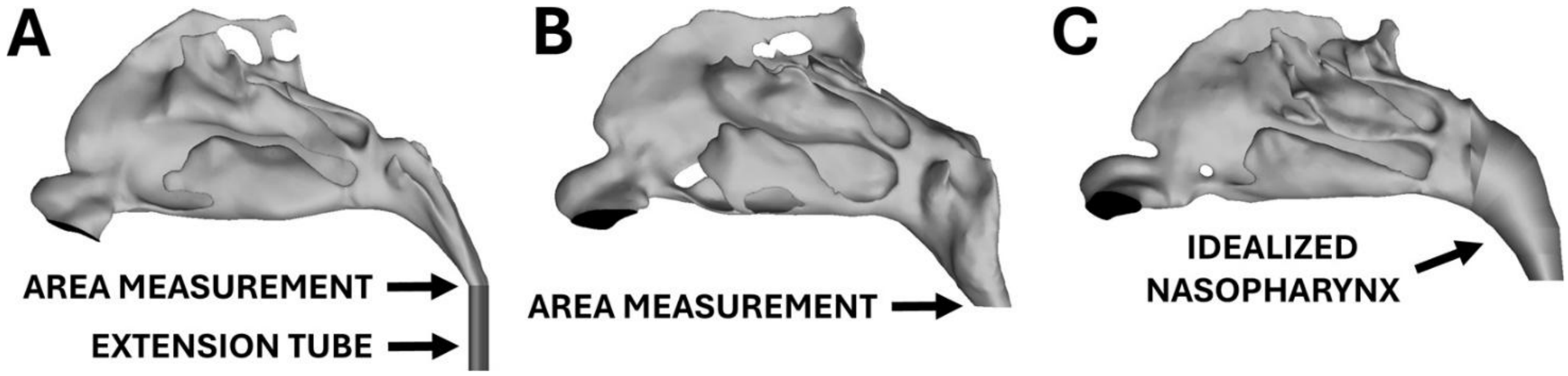

**Fig. 5** The sample of 30 patient-specific models of the nasal cavity, included 19 models with an extension tube (panel A), 4 models without an extension tube (panel B), and 7 models with an idealized nasopharynx (panel C). The accuracy of the CSAs estimated from depth maps was evaluated by measuring the CSA at the distal end of the nasopharynx in 18 models with an anatomically accurate nasopharynx (panel A).

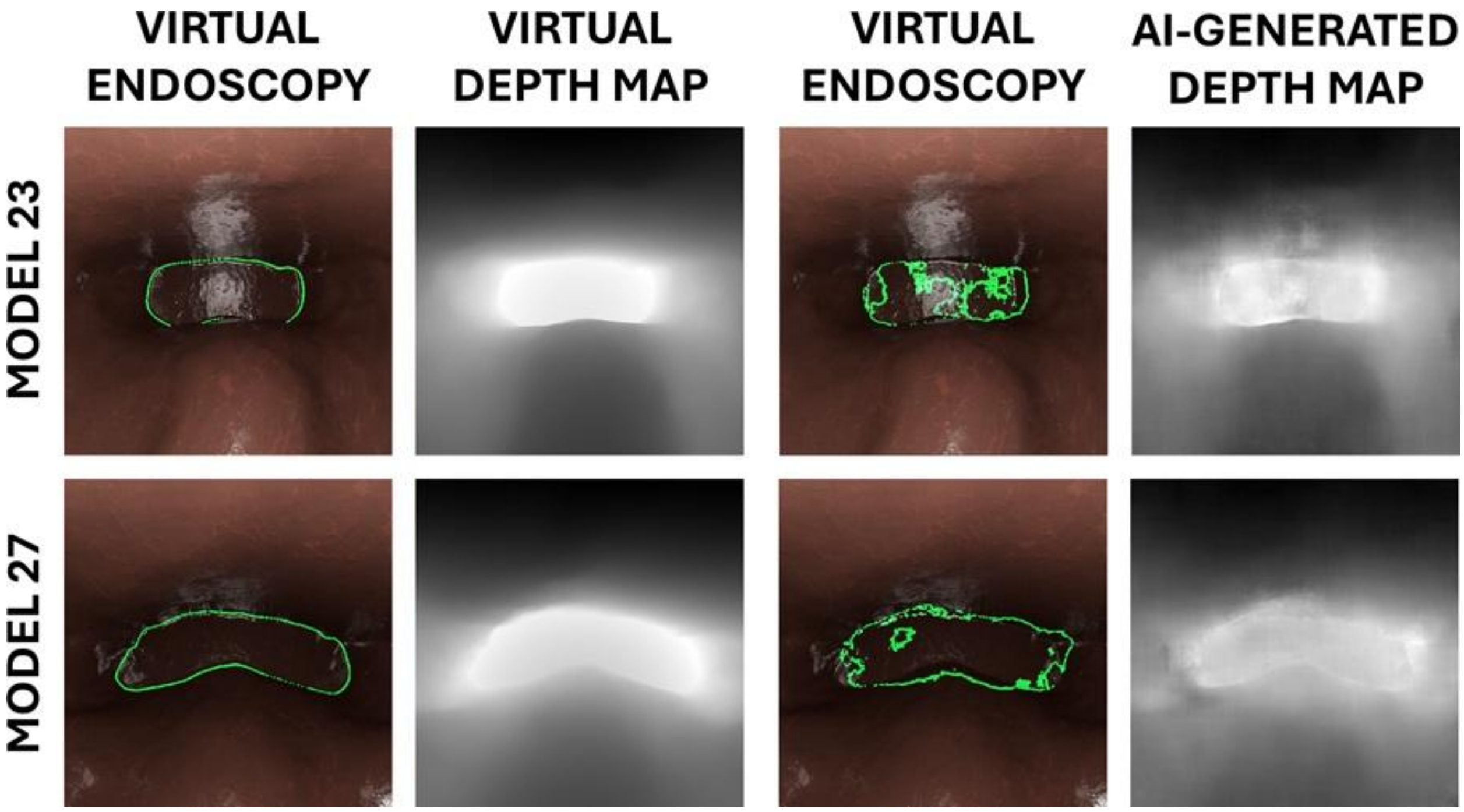

**Fig. 6** Examples of models excluded from the cross-sectional area accuracy analysis. These models lacked an extension tube (**Figure 5B**). Light reflections on the non-anatomical distal wall (corresponding to the distal boundary of the CT segmentation) compromised the accuracy of the AI-generated depth map, resulting in poor delineation of the airway perimeter.

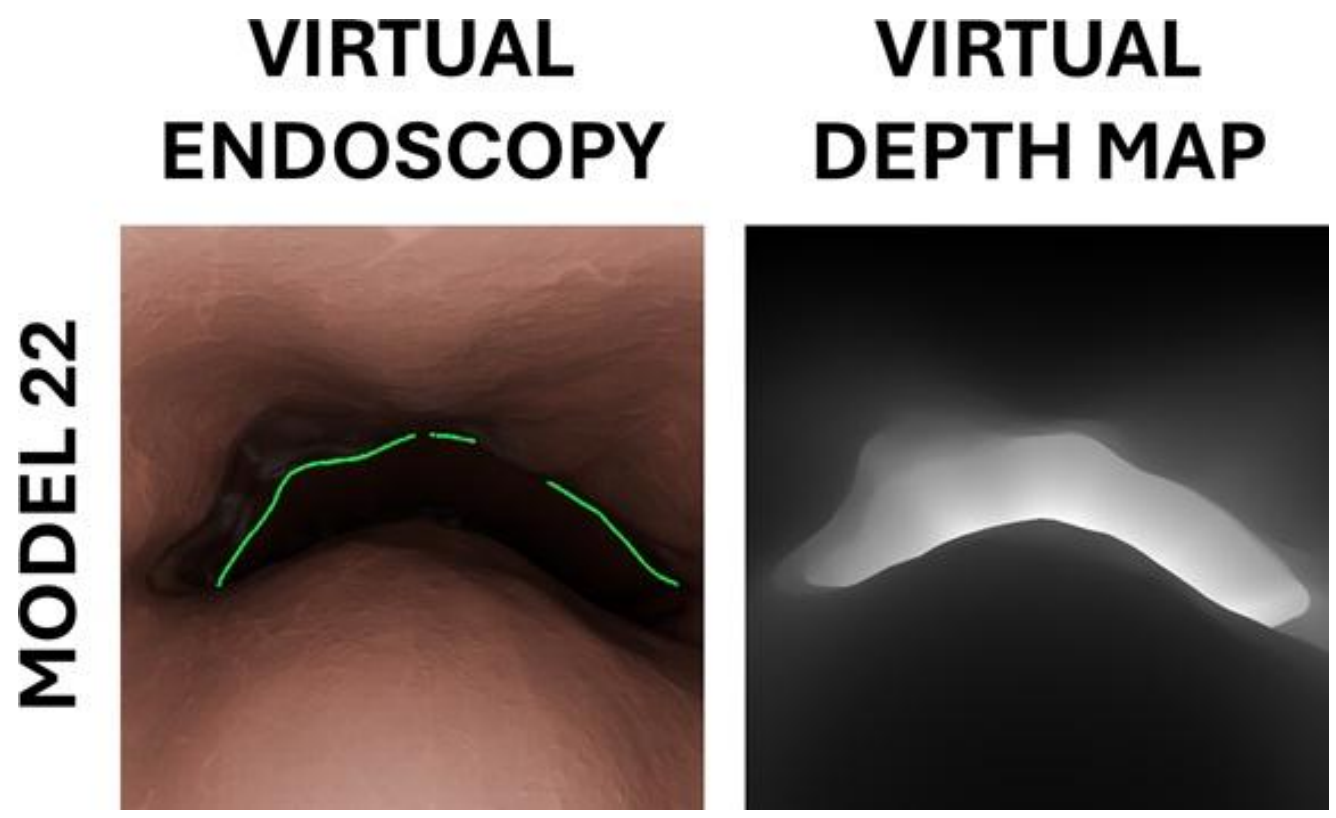

**Fig. 7** Model 22 was excluded from the cross-sectional area accuracy analysis. A hump created by the soft palate partially obstructed the measurement site, preventing visualization of the anterior airway boundary. As a result, the anterior perimeter could not be reconstructed from the point cloud, even when using the ground-truth depth map.

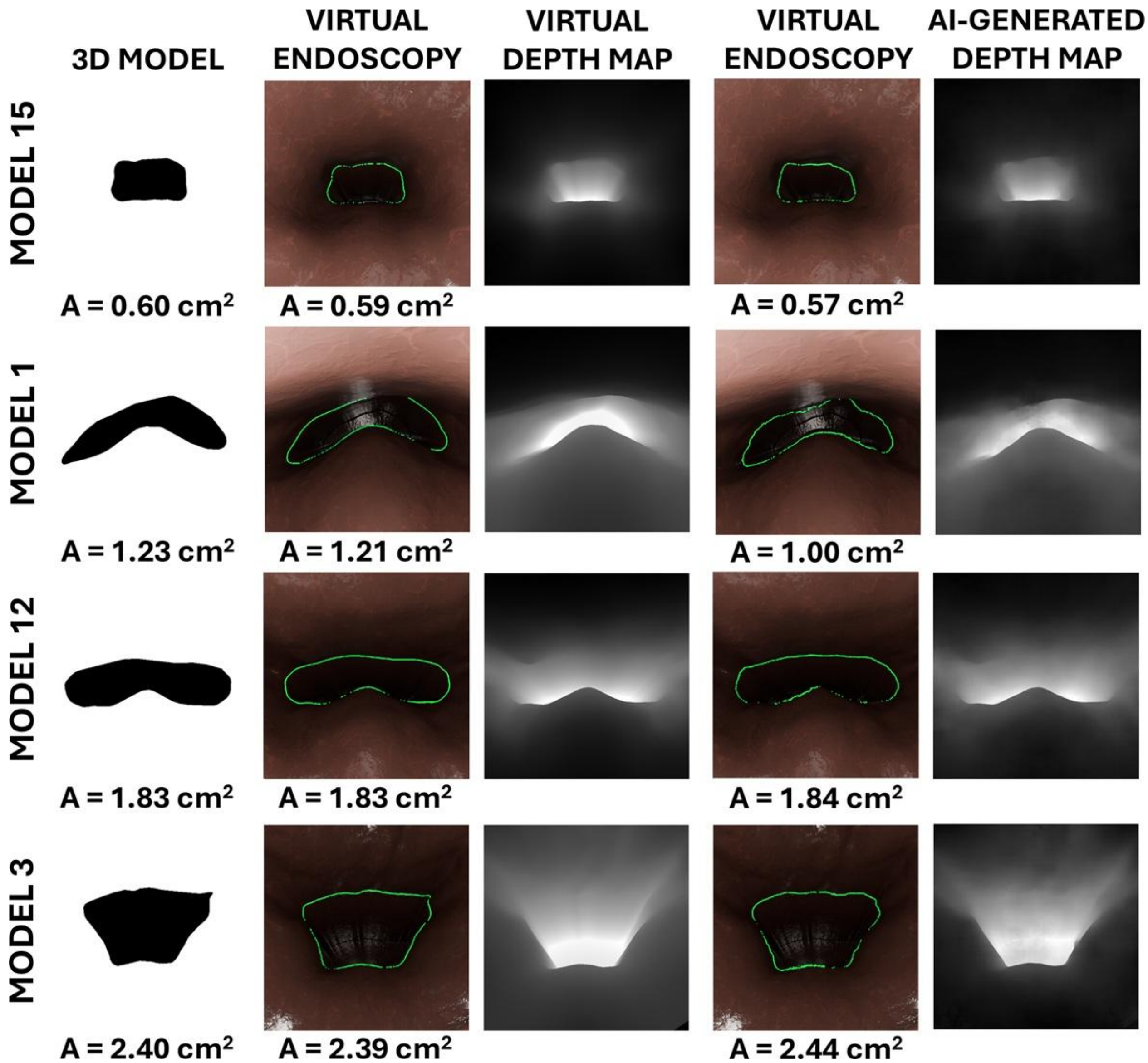

**Fig. 8** Comparison of the planar section at the distal end of the anatomically-accurate nasopharynx in the 3D models with airway perimeter segmentations derived from the ground-truth depth map and AI-generated depth map. Symbol: A = cross-sectional area.

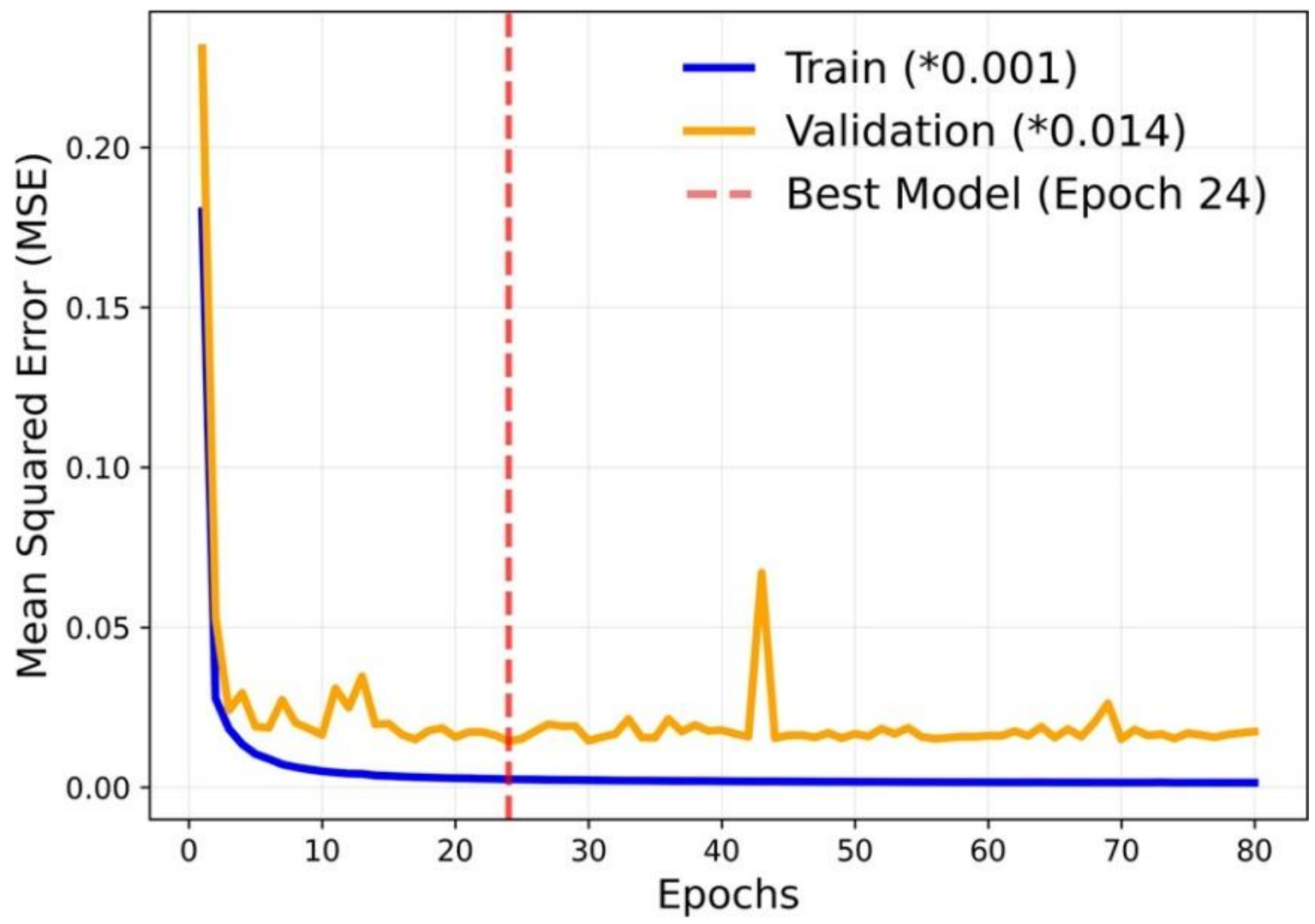

**Fig. 9** Training and validation loss curves. Mean squared error (MSE) for training (blue) and validation (orange) sets across 80 epochs. The best model (red dashed line) was selected at epoch 24, achieving the lowest MSE of 0.014.

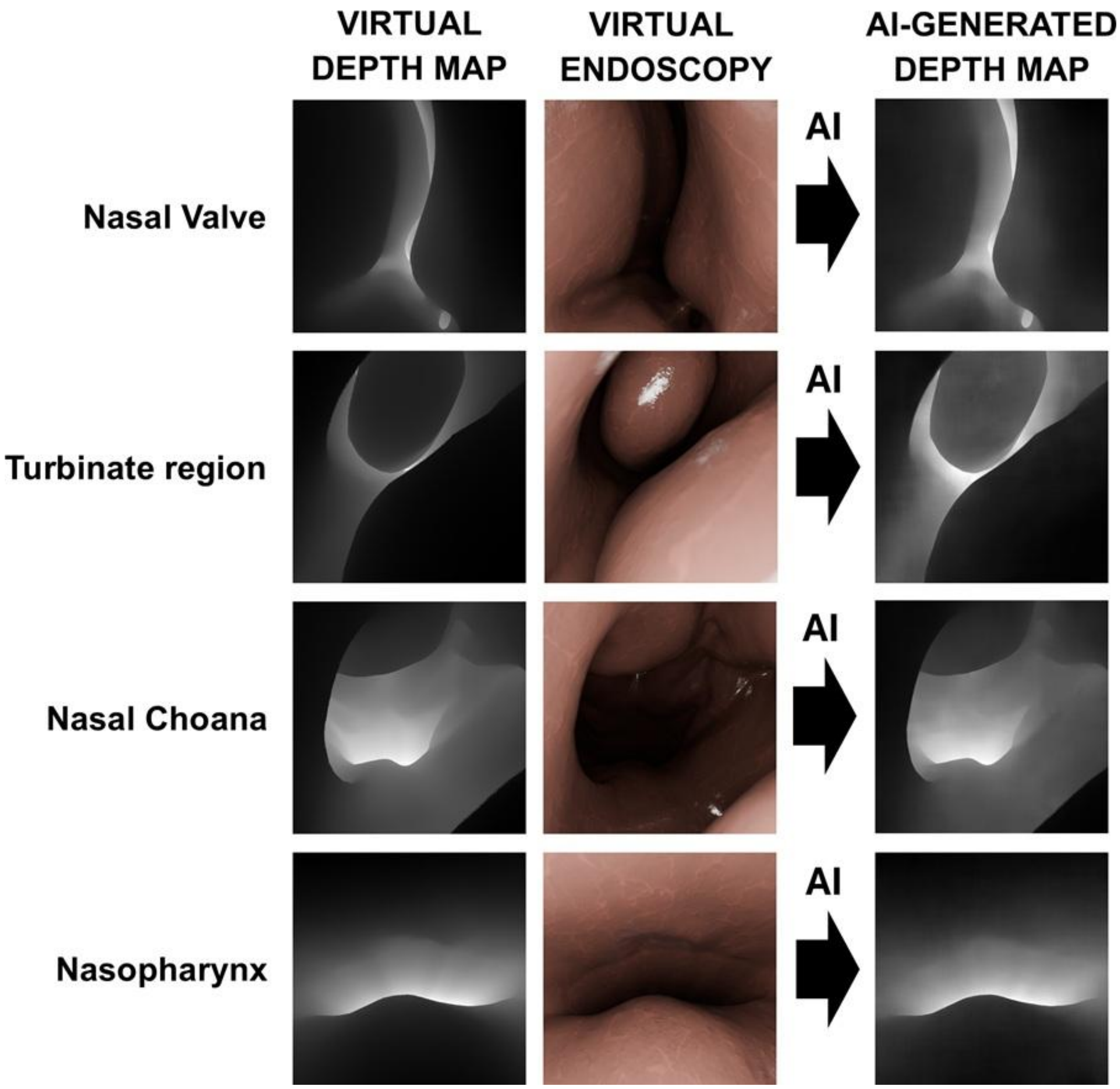

**Fig. 10** Qualitative comparison of AI-generated depth maps with ground-truth depth maps.

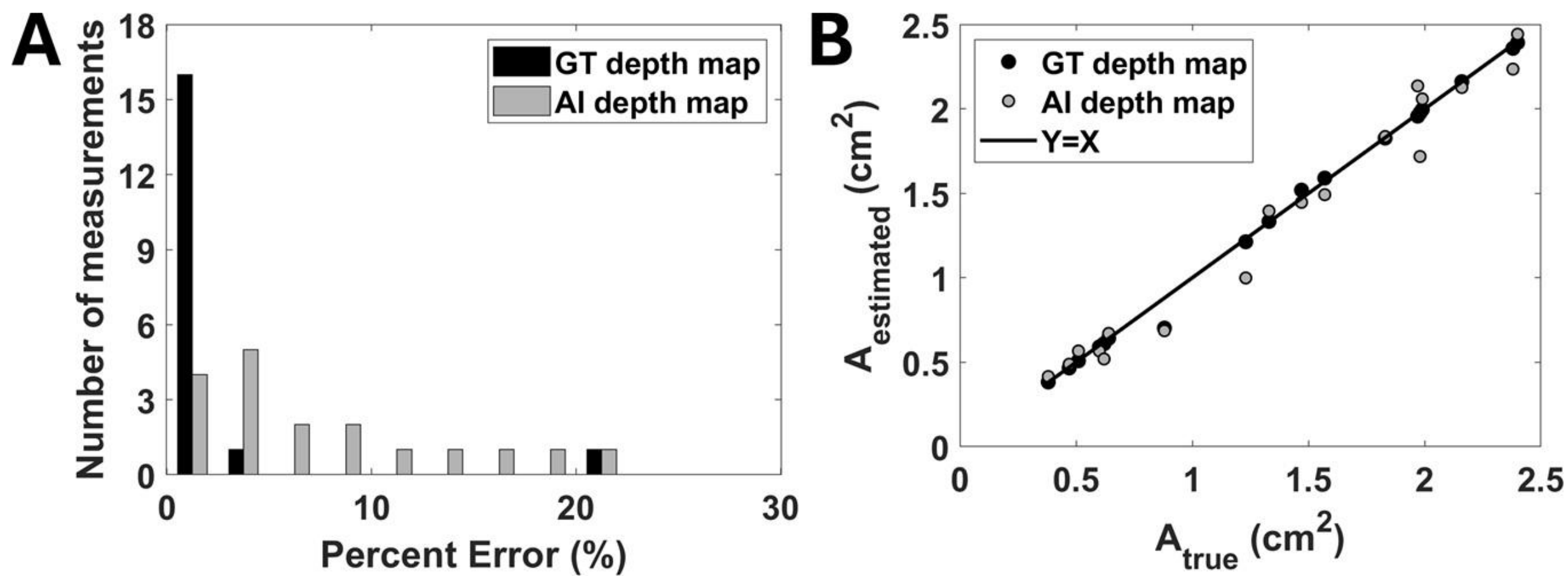

**Fig. 11** (A) Histogram showing the percent error of airspace cross-sectional areas estimated from the depth maps relative to the true areas measured in the 3D models. (B) Scatter plot comparing estimated areas versus true areas measured in the 3D models. Abbreviations: GT = ground truth, AI = artificial intelligence.

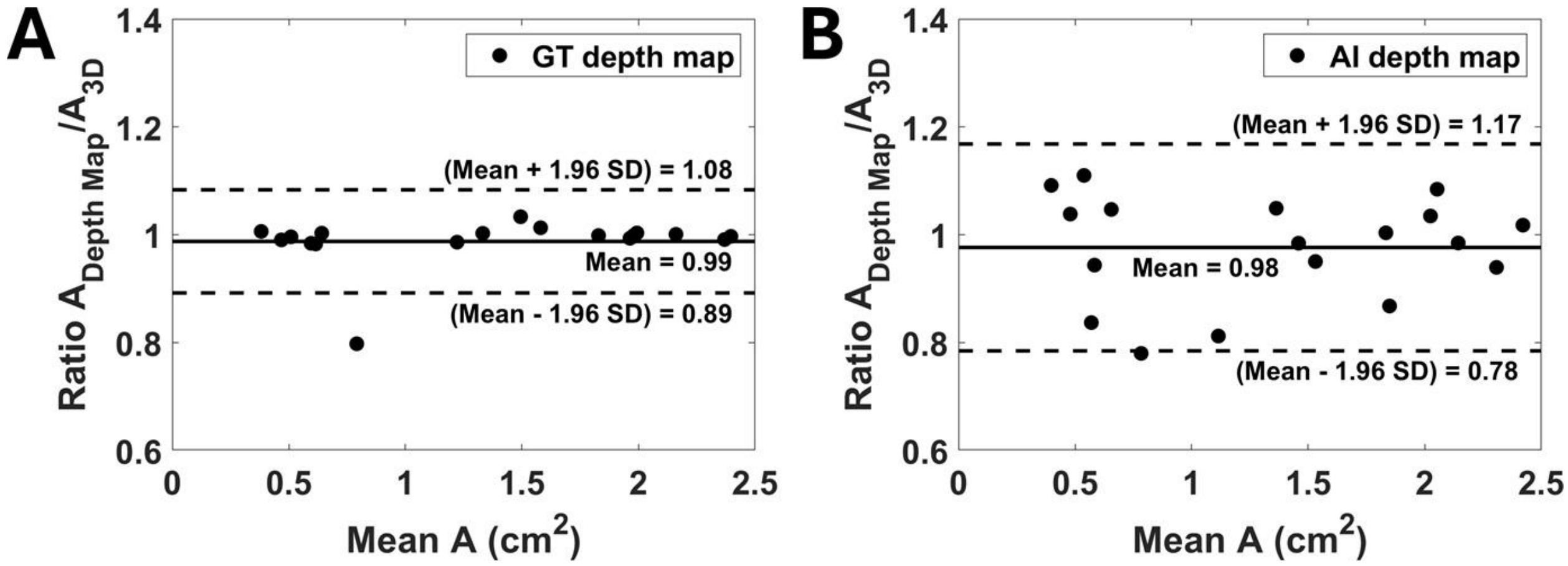

**Fig. 12** Bland-Altman plot for the agreement between airway cross-sectional areas estimated from depth maps and measured in the 3D models. The ratio of the two methods $A_{Depth\ Map}/A_{3D}$ is plotted against the mean of the two methods $(A_{Depth\ Map} + A_{3D})/2$. (A) Ground truth depth map. (B) AI-generated depth map. Abbreviations: GT = ground truth, AI = artificial intelligence.